\documentclass[%
 aip,
 jmp,%
 amsmath,amssymb,
 reprint,%
]{revtex4-1}

\usepackage{graphicx}
\usepackage{dcolumn}
\usepackage{bm}

\begin{document}

\preprint{AIP/123-QED}

\title[Substrate-induced half-metallic property in epitaxial silicene]{Substrate-induced half-metallic property in epitaxial silicene}

\author{Yan Qian}
 \email{qianyan@njust.edu.cn}
\author{Erjun Kan}
\author{Kaiming Deng}
\author{Haiping Wu}%
 \email{mrhpwu@njust.edu.cn}

\affiliation{
Department of Applied Physics, Nanjing University of Science and
Technology, Nanjing 210094, China
}%

\date{\today}

\begin{abstract}
For most practical applications in electronic devices, two-dimensional materials should be transferred onto semiconducting or insulating substrates, since they are usually generated on metallic substrates. However, the transfer often leads to wrinkles, damages, contaminations and so on which would destroy the intrinsic properties of samples. Thus, generating two-dimensional materials directly on nonmetallic substrates has been a desirable goal for a long time. Here, via a swarm structure search method and density functional theory, we employed an insulating N-terminated cubic boron nitride(111) surface as a substrate for the generation of silicene. The result shows that the silicene behaves as a ferromagnetic half-metal because of the strong interaction between silicon and surface nitrogen atoms. The magnetic moments are mainly located on surface nitrogen sites without bonding silicon atoms and the value is $\sim$0.12 $\mu$$_{B}$. In spin-up channel, it behaves as a direct band gap semiconductor with a gap of $\sim$1.35 eV, while it exhibits metallic characteristic in spin-down channel, and the half-metallic band gap is $\sim$0.11 eV. Besides, both the magnetic and electronic properties are not sensitive to the external compressive strain. This work maybe open a way for the utility of silicene in spintronic field.
\end{abstract}

\maketitle

Low-dimensional materials show great important for the development of modern industry nowadays, on account of the demand of micromation the electronic devices.\cite{Boon,Tang,Rao} Fortunately, many of such materials are continuously reported stimulated by the exfoliation of graphene in 2004,\cite{Novoselov1,Novoselov2} e.g. element two-dimensional (2D) materials,\cite{Kong} compound 2D materials,\cite{Mounet} and so on.

For the applications in different fields, the initial properties of the freestanding 2D materials should be modified. Such as graphene and silicene, they both exhibit gapless semimetallic characteristic with Dirac points. But if applied to most practical electronic component, they should possess a large enough band gap. This attracts a great interest in opening an effective band gap of such materials by diverse methods.\cite{Zhou,Du,Ni}. For example, Zhou et al. opened a band gap in graphene by experimentally growing it on SiC substrate.\cite{Zhou} Du et al. tuned silicene from semimetal to semiconductor through the adsorption of oxygen atoms in experiment.\cite{Du} Ni et al. theoretically tuned silicene into a semiconductor by a vertical electric field.\cite{Ni} Nevertheless, most of the reported methods are hard to be widely used in practical applications up to now. This fact encourages some researchers to search for other two-dimensional allotropes. For instance, Luo et al. theoretically designed silicon-based layered structures with semiconducting property,\cite{Luo} Matusalem et al. studied six different 2D silicon allotropes by theoretical calculations, and we also theoretically predicted some monolayered silicon allotropes with diverse properties previously.\cite{Wu,Qian}

On the other hand, via the previous experimental technologies, the most 2D materials are grown on metallic substrates. However, for practical applications, 2D materials must be adhered onto some semiconducting or insulating substrates. This results in that the 2D materials must be transferred to the destination semiconducting or insulating substrates. But this transfer would induce many defects in the samples, such as wrinkles, damages, contaminations and so on, causing the great change in intrinsic properties of the samples. Additionally, the different interaction triggered by the different substrates is another factor in possible modification of their properties. Whereas this interaction in turn can be used as effective means to tune the property of the adhered 2D materials. Thus, instead of regulating the property of 2D materials via the mentioned methods, searching for a suitably nonmetallic substrate has been a promising way for property engineering, and a few literatures have been reported as the following. Graphene was experimentally grown on nonmetallic substrates like SiC,\cite{Zhou} sapphire,\cite{Song} and SrTiO3.\cite{Sun} SiC substrate can induce a band gap of $\sim$0.26 eV in epitaxial graphene,\cite{Zhou}  graphene on sapphire or SrTiO3 substrates could be used to fabricate a high performance field-effect transistor (FET).\cite{Song,Sun} Graphene was also grown on 2D materials in experiment, such as hexagonal boron nitride (h-BN).\cite{Son} For silicene, very few literatures were reported on its generation on nonmetallic substrates.\cite{Chen,Houssa,Chiappe} In some theoretical works, Chen et al. generated silicen on Al$_{2}$O$_{3}$,\cite{Chen} and Houssa et al. chose ZnS as a substrate.\cite{Houssa} Besides, Chiappe et al. experimentally reported on the epitaxy of silicon nanosheet on MoS$_{2}$ subatrate.\cite{Chiappe} Notably, these 2D materials still need great improvement, because they are mostly defective or nonuniform in experiment.

Excitedly, just in recent, large-scale of high-quality graphene was successfully grown on non-metallic substrates by experimental method.\cite{Park,Kolmer}  This experimental work encourage us to continuously explore the generation of 2D materials on nonmetallic substrates within the framework of the density functional theory, which is expected to give a guide for experimental synthesis. Owing to the fact that most of the previous electronic components are made of silicon-based materials, and 2D silicon materials are the only ones that can match well with them, the research content in this work is focused on the generation of 2D silicon on a nonmetallic substrate (i.e. cubic boron nitride (c-BN)). The calculated result shows that silicene generated on N-terminated c-BN(111) surface behaves as ferromagnetic half-metal and is unsensitive to the external compression.

Here, the approach is outlined in Fig. 1a. It is typically divided into three regions: the bulk region, the unreconstructed surface, and the
reconstructed surface. The bottom side of the slab is terminated by hydrogen atoms. The repeated slab geometry is separated by a vacuum
region of 25 {\AA}. The atoms in the bulk region of the slab model are fixed in order to remain the bulk nature of the material, the atoms in the
unreconstructed surface region are relaxable but not plunged into the structure evolution, and the atoms in the reconstructed surface
region are subject to swarm-structure evolution. The generation of initial structures and swarm-structure evolution are implemented in CALYPSO code,\cite{Wang1,Wang2} and more details can refer to the previous work\cite{Lu,Wu}. Once the initial structures are generated, the geometries will be optimized
\begin{figure}[htbp]
\centering
\includegraphics[width=8.5cm]{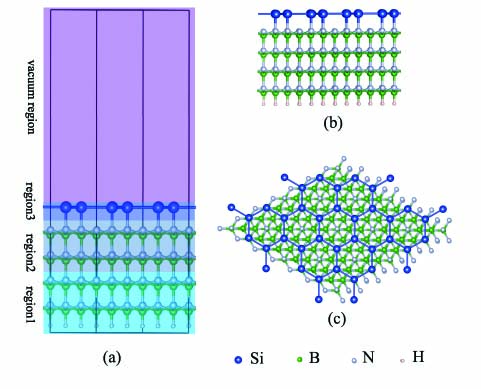}
\caption{(Color online). (a) is a sketch of the slab model. Si atoms are deposited on the ideal N-terminated c-BN(111) 3$\times$3 surface. The region1 is the bulk region, the region2 is the unreconstructed surface, and the region3 is the reconstructed surface. (b) and (c) are the side and top views of the model.}
\label{fig:Figure1}
\end{figure}
nextly. The underlying ab initio structural relaxations and electronic band structure calculations are carried out in the framework of density functional theory (DFT) within generalized-gradient approximations using the PerdewBurke-Ernzerhof (PBE)\cite{Perdew} exchange correlation functional and projector-augmented-wave (PAW)\cite{Kresse1} potentials as performed in VASP.\cite{Kresse2} To ensure high accuracy, the k-point density and the plane waves cutoff energy are increased until the change in the total energy is less than 10$^{-5}$ eV, and the Brillouin-zone (BZ) integration is carried out using 11$\times$11$\times$1 Monkhorstack grid in the first BZ, with plane waves of kinetic energy up to 600 eV being employed. Structural relaxations are performed until the Hellmann-Feynman force on each atom is reduced by less than 0.001 eV{\AA}$^{-1}$. The adopted PAW pseudopotentials of B, N, and Si treat the 2s$^{2}$2p$^{1}$, 2s$^{2}$2p$^{3}$, and 3s$^{2}$3p$^{2}$ electrons as valence electrons.

In order to confirm the ground-state magnetic phase of two-dimensional silicon material after getting the evolutive structure generated on N-terminated c-BN(111) 3$\times$3 surface, the total energies of nonmagnetic and ferromagnetic phases are calculated. The result tells that the ferromagnetic structure is energetically stable than the nonmagnetic one by $\sim$ 0.06 eV per/unicell. Thus, the properties of ferromagnetic phase is discussed in the following if not specially denoted. For the structural properties, the adhered monolayered silicon displays hexagonal lattice as pictured in Figs. 1b and 1c, but it exhibits absolute plane structure, instead of buckling property as in free-standing silicene. This absolutely plane structure is the same as those of the other monolayered silicon materials generated on the B-terminated c-BN(111) surface\cite{Wu}, and the performance is driven by the strong interaction between Si and surface N atoms. The electron localization functions (ELF) plotted in Fig. 2a can clearly confirm this strong interaction, there exist strong $\sigma$
\begin{figure}[htbp]
\centering
\includegraphics[width=8.5cm]{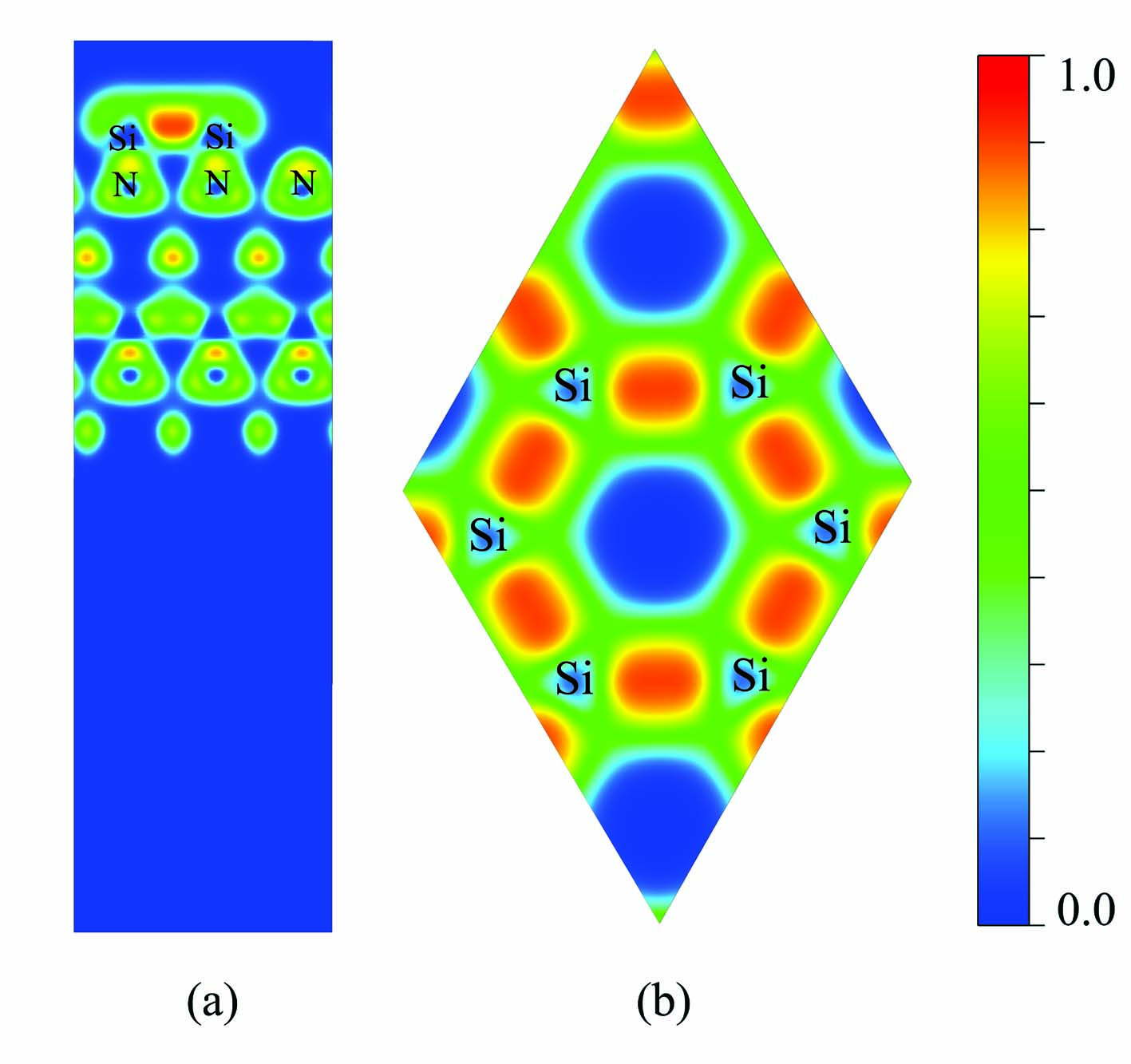}
\caption{(Color online). The calculated electron localization functions (ELF). (a) and (b) are for (001) and (100) planes, respectively.}
\label{fig:Figure2}
\end{figure}
bonds between Si and surface N atoms. The length of Si-N bonds, i.e. the distance between the Si monolayer and substrate surface, is $\sim$1.83 {\AA}, this value is greatly shorter than 1.95 {\AA} between the silicon monolayers and the B-terminated c-BN(111) surface\cite{Wu}. This performance can be explained by the shorter radius of N atom and stronger coupling between Si and surface N atoms. However, this value is larger than $\sim$1.73 {\AA} in Si$_{3}$N$_{4}$ and 1.76 {\AA} in 2D SiN given by calculations or experiments\cite{QianWu,Hardie}, the reason is that the formation of B-N bonds weakens the interaction between Si and surface N atoms in this sample. The length of Si-Si bonds is $\sim$2.56 {\AA}, this value is some larger than 2.37 and 2.28 {\AA} in monocrystalline Si and silicene calculated in this work via the same accuracy. This change is originated from the hardness of c-BN substrate and the strong interaction between Si and N atoms. Since c-BN is hard to be compressed or stretched, and associated with the strong interaction between Si and N atoms, the Si-Si bonds are stretched largely compared with the other Si allotropes. Although the Si-Si bonds are lengthened, there still exists strong interaction between the two neighboring silicon atoms as in Fig. 2b. The figure expressly shows that strong $\sigma$ bonds are formed between the two neighboring Si atoms.

Furthermore, in order to clarify the energetics of the silicon monolayers generated on the N-terminated c-BN(111) surface, the cohesive energy \textit{E}$_{c}$ and binding energy \textit{E}$_{b}$ are also calculated. \textit{E}$_{c}$ and \textit{E}$_{b}$ are defined as:

\textit{E}$_{c}$=(\textit{E}$_{subs}$+\textit{n}$_{Si}$$\mu$$_{Si}$-\textit{E}$_{tot}$))/\textit{n}$_{Si}$

\textit{E}$_{b}$=(\textit{E}$_{subs}$+\textit{E}$_{Si-monolayer}$-\textit{E}$_{tot}$)/\textit{n}$_{Si}$

Here, {E}$_{tot}$, {E}$_{subs}$ and {E}$_{Si-monolayer}$ are the total energies of the reconstructed structure, the c-BN substrate and the freestanding
Si monolayer, respectively, {n}$_{Si}$ is the number of Si atoms in the Si monolayer, and ${\mu}$$_{Si}$ is the chemical potential of Si. This definition notes that the binding energy is the energy gain per Si atom in the deposition of the Si monolayer on the substrate. When ${\mu}$$_{Si}$ is takes as the total energy of an isolated Si atom, then the cohesive energy means the energy gain to make the Si monolayer on the substrate from the Si atoms. The calculated result indicates that the \textit{E}$_{c}$ and \textit{E}$_{b}$ are 6.94 and 2.17 eV, revealing that the structure is thermodynamically stable with positive binding and cohesive energies. In addition, \textit{E}$_{c}$ of 6.94 eV is larger than 5.41 and 4.78 eV for those of monocrystalline silicon and freestanding silicene (both values are from our PBE calculations), implying that the c-BN(111) surface is suitable for the generation of silicon monolayers. This value is also larger than those of the other silicon allotropes generated on the B-terminated c-BN(111) surface\cite{Wu}, arisen from the fact that the interaction between Si and N atoms is stronger than the one between Si and B atoms.

Based on the possibility of generating Si monolayers on the N-terminated c-BN(111) surface, the other properties of the sample were explored furthermore. Fig. 3 draws the total density of electronic states (DOS) and band structure of the sample. From the DOS in Fig. 3a, it unveils that the sample behaves as a semiconductor with a band gap of $\sim$1.35 eV in spin-up channel, while it exhibits metallic character with many electronic states located at the Fermi energy level (\textit{E}$_{F}$) in spin-down channel. This feature manifests that the sample is a half-metal and possesses a half-metallic gap (the minimal energy gap for a spin excitation) of $\sim$0.11 eV. The band structure plotted in Fig. 3b reveals that it is a direct band gap semiconductor in spin-up channel, since the conduction band minimum (CBM) and the valence band maximum (VBM) are both situated at the $\Gamma$ reciprocal
\begin{figure}[htbp]
\centering
\includegraphics[width=8.5cm]{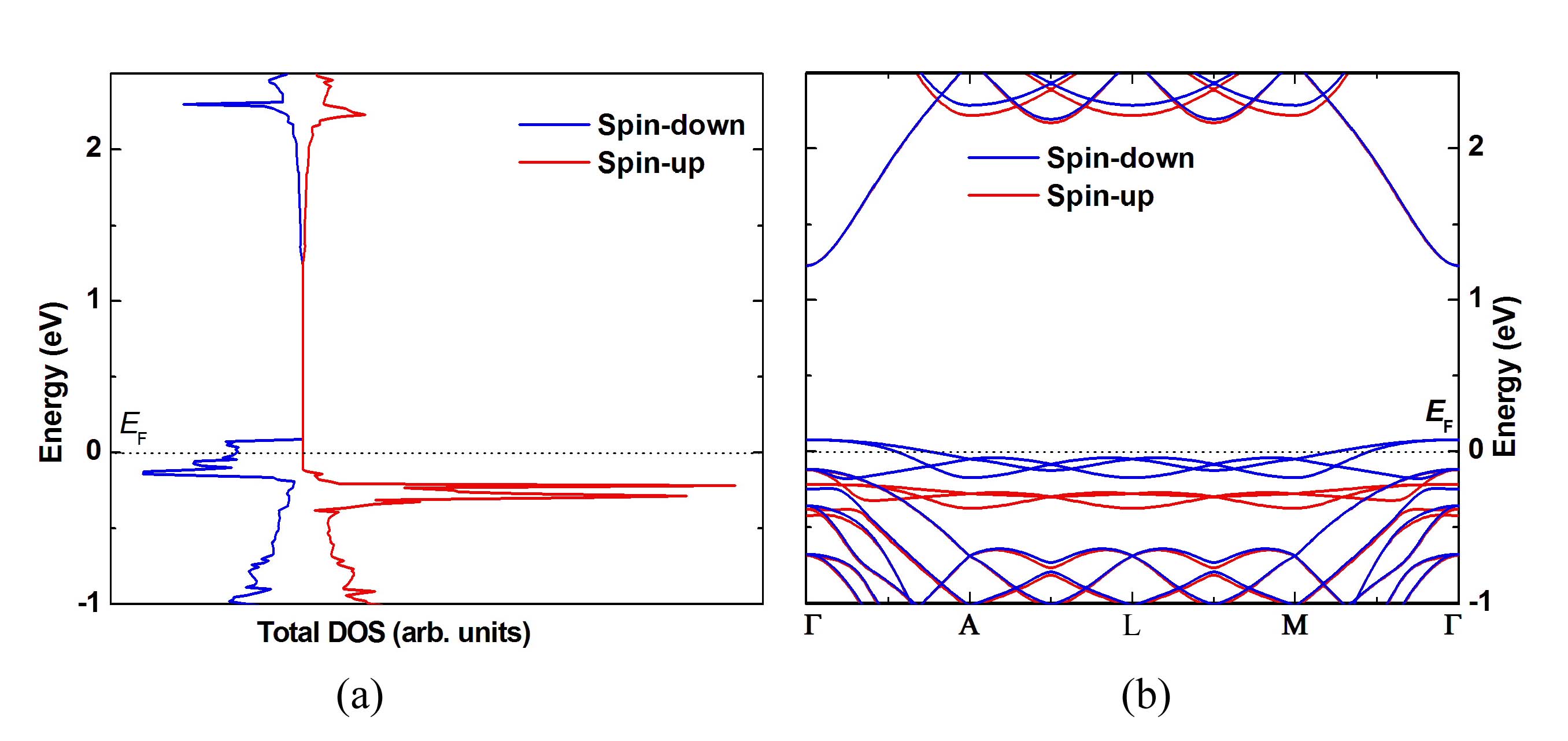}
\caption{(Color online). (a) and (b) are the density of electronic states (DOS) and band structure for the sample, respectively.}
\label{fig:Figure3}
\end{figure}
point. The partial density of electronic states (PDOS) near \textit{E}$_{F}$ is pictured in Fig. 4 for more detail. In both spin channels, Fig. 4a notes that the valence bands are mainly constituted by the states of the substrate, and a few are originated from Si 3p states, while the conduction bands are mainly composed of Si 3p states and scarcely from the states of the substrate. The PDOS of N and B atoms located on different sites as shown in Fig. 4b tells that the states at \textit{E}$_{F}$ are majorally from 2p orbitals of the surface N atoms which are not bonded with Si atoms, few states are contributed by the other atoms in the first top BN layer. This fact elucidates that the carriers can transfer only along the silicon monolayer and the first top BN layer on c-BN(111) surface, and the deeper layers of the substrate still remain semiconducting characteristic.
\begin{figure}[htbp]
\centering
\includegraphics[width=8.5cm]{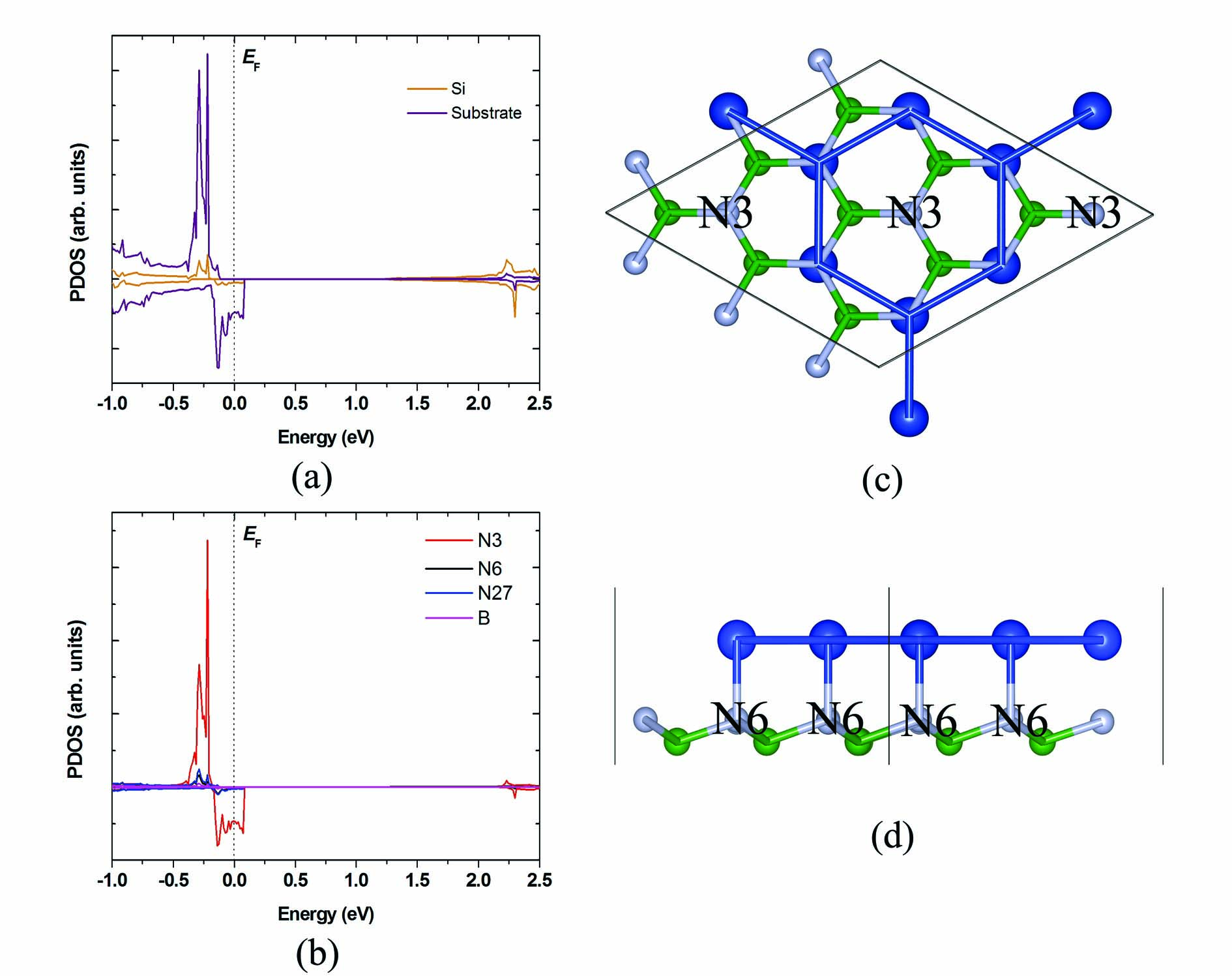}
\caption{(Color online). (a) The PDOS of silicene and substrate. (b) The PDOS of N and B atoms stated on different sites. (c) and (d) are the top and sides views for silicene and the first BN layer of the substrate. N3 represents the surface N atoms not bonded with Si atoms, N6 is for the ones bonded with Si atoms, N27 represents the other N atoms of the substrate.}
\label{fig:Figure4}
\end{figure}

Nextly, the magnetic property is discussed here. The magnetic moments are mostly situated at the surface N sites not bonded with Si atoms, and the value is $\sim$0.12 $\mu$$_{B}$. Besides, there are $\sim$0.01 $\mu$$_{B}$ located on the Si sites. This distribution of magnetic moments is caused by the unsaturated orbitals of the surface N atoms without bonding with Si atoms, while the orbitals of the other N atoms are all bonding with the neighboring atoms.

Since the electronic components usually work under external pressure, we also studied the effect of external compressive strain on the properties of the sample. Figure 5 pictures the changing tendency of magnetic and electronic properties of the sample when the external compressive strain perpendicular to the surface is applied. The result tells that the magnetic moments still keep around 0.12 and 0.01 $\mu$$_{B}$ on N (without bonding Si atoms) and Si sites, and the half-metallic nature is well kept in the range of the strain employed in this work. This fact demonstrates that the properties of the sample are unsensitive to the external compressive strain.
\begin{figure}[htbp]
\centering
\includegraphics[width=9cm]{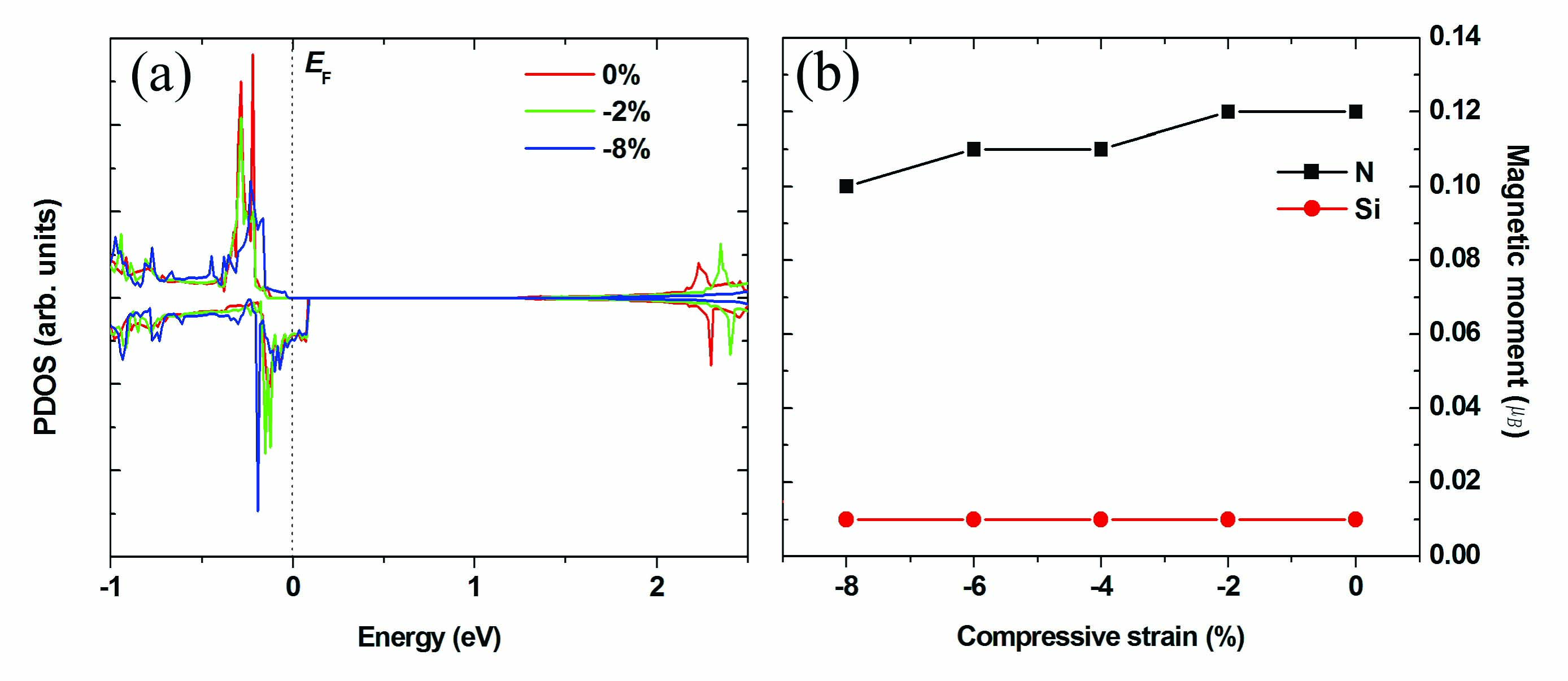}
\caption{(Color online). (a) The DOS of the material at the external strain of -2\% and -8\% as representative. (b) The dependence of magnetic  on the external strain.}
\label{fig:Figure5}
\end{figure}

Finally, the electronic property of the nonmagnetic material is studied as well, and the DOS and band structure are plotted in Fig. 6. From Fig. 6a, it displays that the sample exhibits semiconducting nature with a gap of $\sim$ 1.32 eV, and this gap is mostly equal to that of the half-metallic one in spin-up channel. Fig. 6b deeply shows that CBM and VBM are both located at the $\Gamma$ point, illustrating that it is a direct band gap semiconductor.
\begin{figure}[htbp]
\centering
\includegraphics[width=9cm]{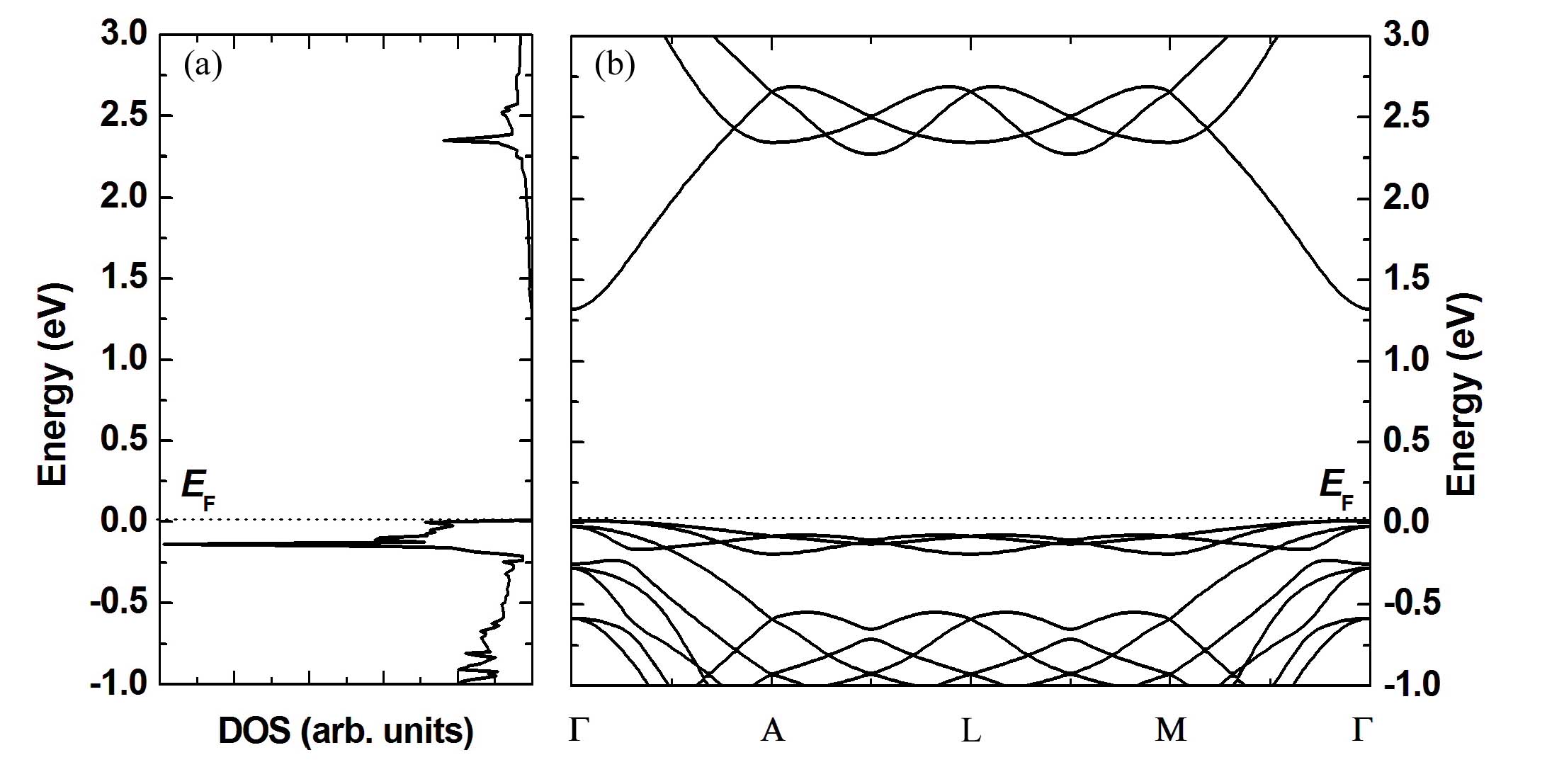}
\caption{(Color online). (a) The DOS and band structure of the nonmagnetic one.}
\label{fig:Figure6}
\end{figure}

In conclusion, a ferromagnetic silicene with half-metallic nature is found when it is generated on N-terminated c-BN(111) surface. The half-metallic band gap is $\sim$0.11 eV, and the magnetic moments are mainly located at the surface N atoms not bonded with Si atoms. These properties are not sensitive to the external compressive strain. The nonmagnetic one shows semiconducting property with a gap of $\sim$ 1.32 eV. This work gives a promising direction of applying, engineering or searching for two-dimensional silicon materials with diverse properties.

\vspace{1ex}

This work was supported by the National Natural Science Foundation of China (Grant nos. 11404168, 11304155, and 11374160).

\end{document}